\documentclass[aps,letterpaper,twocolumn,preprintnumbers,floatfix,superscriptaddress,nofootinbib]{revtex4-1}

\usepackage{amsmath}
\usepackage{amsfonts}
\usepackage{amssymb}
\usepackage{graphicx, rotating}
\usepackage{epstopdf}
\usepackage{epsfig}
\usepackage{latexsym}
\usepackage{color}
\usepackage[dvipsnames]{xcolor}
\usepackage{multirow}

\usepackage{xcolor,colortbl}

\usepackage{slashed}
\usepackage{hyperref}
\hypersetup{colorlinks, citecolor=bluscuro, linkcolor=bluscuro, urlcolor=bluscuro}
\definecolor{rossos}{cmyk}{0,1,1,0.55}
\definecolor{bluscuro}{rgb}{0.15, 0.2, .85}
\definecolor{bluchiaro}{cmyk}{1,.3,0.,0.1}
\definecolor{verdescuro}{rgb}{0.3,0.8,0.3}

\newcommand{\eq}[1]{Eq.~(\ref{#1})}

\newcommand{\nn}{\nonumber}

\newcommand{\be}{\begin{equation}}
\newcommand{\ee}{\end{equation}}          
\newcommand{\bea}{\begin{eqnarray}}
\newcommand{\eea}{\end{eqnarray}}

\newcommand{\M}{{\cal M}}
\newcommand{\A}{{\cal A}}

  \def\mpl{m_\mathrm{Pl}}

\begin{document}

\widetext

\begin{flushright}
{\small 
Saclay-t23/017}
\end{flushright}

\title{Massive Gravity is not Positive}

\author{Brando Bellazzini}
\affiliation{Universit\'e Paris-Saclay, CNRS, CEA, Institut de Physique Th\'eorique, 91191, Gif-sur-Yvette, France. }
\author{Giulia Isabella}
\affiliation{Universit\'e Paris-Saclay, CNRS, CEA, Institut de Physique Th\'eorique, 91191, Gif-sur-Yvette, France. }
\affiliation{D\'epartment de Physique Th\'eorique, Universit\'e de Gen\`eve, 24 quai Ernest-Ansermet, 1211 Gen\`eve 4, Switzerland}
\affiliation{ Universit\'e Paris-Saclay, CNRS/IN2P3, IJCLab, 91405 Orsay}
\author{Sara Ricossa}
\author{Francesco Riva}
\affiliation{D\'epartment de Physique Th\'eorique, Universit\'e de Gen\`eve, 24 quai Ernest-Ansermet, 1211 Gen\`eve 4, Switzerland}

\begin{abstract}
\noindent 

We derive new positivity bounds at finite momentum transfer, assuming a large separation between the mass  $m$ of the lightest particle in the effective theory  and the mass gap $M$ to new heavy states.  Massive gravity parametrically violates these bounds unless the cutoff is  within one order of magnitude of the graviton mass   $M\lesssim O(10)m$. Non-gravitational effective theories of massive spin-2 particles are similarly bounded.

 \end{abstract}

\maketitle
\medskip

%%%%%%%%%%%%%%%%%%%%%%%%%%%%%%%%%%%%%%%%%%%%%%%%%%%%%%%%%%%%%%%

\section{Introduction}

The principles of causality, unitarity, crossing symmetry, and Lorentz invariance  enforce non-trivial constraints on otherwise healthy-looking effective field theories (EFT), which describe the emergent infrared (IR) dynamics relevant to low energy  observers. 
The simplest of these constraints  take the form of inequalities among scattering amplitudes or Wilson coefficients, and are therefore known as \emph{positivity bounds}.
These  have found interesting applications in particle physics and cosmology, allowing to discern  EFTs that can have  a consistent ultraviolet (UV) description---the EFT landscape---from theories that do not---the EFT swampland.

Positivity bounds shape the space of EFT amplitudes, by constraining the structure of  higher derivative interactions. These have played a particularly important role in our understanding of possible departures from Einstein gravity,   see e.g. \cite{Adams:2006sv,Camanho:2014apa,Bellazzini:2015cra,Cheung:2016wjt,Hamada:2018dde,Bonifacio_2018,Bellazzini:2019xts,Melville:2019tdc,Tokuda_2020,Caron-Huot:2021rmr,Bern:2021ppb,Caron-Huot:2021enk,Arkani-Hamed:2021ajd,Bellazzini:2021shn,Caron-Huot:2022ugt,Caron-Huot:2022jli,Chiang:2022jep,Bellazzini:2022wzv,Henriksson:2022oeu,Herrero-Valea:2020wxz,Herrero-Valea:2022lfd,Edelstein_2021,Gonz_lez_2022,Serra:2022pzl,noumi2022finite}.

In this article, we turn our attention towards the theory of dRGT \emph{massive} gravity \cite{Arkani-Hamed:2002bjr,deRham:2010ik,deRham:2010kj}. This is a (generally covariant) EFT of a single massive spin-2 particle, whose   mass~$m$ is experimentally constrained to be of the order of the smallest energy scale in our universe,  the Hubble parameter: $m\sim H_0$ (see e.g.  \cite{deRham:2016nuf} and references therein). 
Our goal is to understand whether this EFT can be used to describe physics at \emph{parametrically} larger scales (distances much shorter than the Hubble radius $H_0^{-1}$), as relevant for any practical and cosmological application.

  Positivity  has already been employed in this context~\cite{Cheung:2016yqr,Bellazzini:2017fep,deRham:2018qqo}. Bounds  restricted to forward-scattering  imply 
   that the two  free parameters of dRGT massive gravity, $c_3$ and $d_5$,  must live in a certain finite compact region \cite{Cheung:2016yqr}, and that the ultimate energy-cutoff $M$ of the theory is  smaller than $M\ll (m^3\mpl)^{1/4} $~\cite{Bellazzini:2016xrt,Bellazzini:2017fep}, with $\mpl$  the reduced Planck mass.\footnote{\label{ftnt1} Since $M$ is much smaller than the  strong coupling scale for longitudinal polarizations $\Lambda_3\equiv(m^2\mpl)^{1/3}$ in dRGT~\cite{Arkani-Hamed:2002bjr}, the new degrees of freedom must UV complete massive gravity in the weakly coupled regime. }   
This scale corresponds to distances $M^{-1}$ roughly larger than   the size of our solar system. 

In this work we extend massive gravity positivity bounds to regimes of large momentum transfer,~$|t|\gg m^2$, exploiting two basic observations.
 Firstly, inelastic matrix elements are bounded by  elastic ones; so finite-$t$ dispersive integrals must be smaller than forward ones.  
 Secondly, for the unknown part of dispersive integrals with c.o.m. energy squared $s\geq M^2\gg|t|$, crossing symmetry is simple and resembles  near-forward or massless scattering.

The  positivity bounds emerging from this analysis lead to a much stronger condition on the regime of validity  of dRGT.
We find that the cutoff of massive gravity is  parametrically close to its mass, and tied to it by the linear relation,
\begin{equation}\label{bound}
M\lesssim O(10)\,\, m  \,,
\end{equation}
independently of all the other parameters in the theory. 
This conclusion can not be avoided by the mechanism of Vainshtein screening \cite{Vainshtein:1972sx,Deffayet:2001uk,deRham:2014zqa}, as all Vainshtein radii from compact sources are smaller than the size of the universe.

 In section \ref{sec:Positiveness} we lay out our assumptions, and derive a simple version of positivity constraints that is suitable to study massive higher-spin scattering. We apply these bounds  to dRGT in section~\ref{sec:positivitydRGT}. In section~\ref{secBDRGT} we show that the bounds apply also to general deformations  from dRGT with higher-derivative interactions, detuned potentials,  and non-gravitational theories.   Finally in section~\ref{sec:conclusions}  we summarize our results and discuss future directions.

\section{Positivity}
\label{sec:Positiveness}

We study the $2\to 2$ scattering of massive gravitons in flat spacetime. 
In what follows we list our assumptions and show how they can be efficiently employed to obtain parametric bounds of the form of \eq{bound}.  \\

\noindent
{\it i) Unitarity of the  $S$-matrix,} \\
\begin{equation}  \label{optical}
\frac{\M-\M^\dagger}{i}=\M^\dagger \M \succeq 0  \,,
\end{equation}
 for physical energies $s\geq 4m^2$.  This equation, evaluated on any complete set of states, is ultimately responsible for positivity. In practical applications only truncated sets of states can be considered (e.g. finite number of partial waves, states of definite helicity, etc\ldots), and each of  these sets  accesses different partial information.

{We work with generic  initial $|1^{\lambda_1} 2^{\lambda_2}\rangle$ and final $|3^{\lambda_3} 4^{\lambda_4}\rangle$ 2-particle states of arbitrary momentum and helicity $\lambda_i$. Here $|3^{\lambda_3} 4^{\lambda_4}\rangle \equiv R(\theta)|1^{\lambda_3} 2^{\lambda_4}\rangle$ is defined by a rotation $R(\theta)=\mathrm{exp}(-iJ_2\theta)$ of an initial state with given helicity.
Positivity of $|\M(|1^{\lambda_1} 2^{\lambda_2}\rangle +e^{i\alpha }  |3^{\lambda_3} 4^{\lambda_4}\rangle )|^2$ for all~$\alpha$, implies,
 \begin{eqnarray}
 \label{hankelcondition}
\! 2\left|\langle 3^{\lambda_3} 4^{\lambda_4} | \M^\dagger\M|1^{\lambda_1}2^{\lambda_2}\rangle\right| &\leq &\\
 \langle 1^{\lambda_1} 2^{\lambda_2} | \M^\dagger\M|1^{\lambda_1} 2^{\lambda_2}\rangle &+&\langle 3^{\lambda_3} 4^{\lambda_4} | \M^\dagger\M|3^{\lambda_3} 4^{\lambda_4}\rangle\,. \nonumber
 \end{eqnarray}
This has a simple, but powerful, physical interpretation: \emph{inelastic  $ \M^\dagger\M$ matrix elements must be smaller than elastic ones.}  By \eq{optical} the same statement holds for $(\M-\M^\dagger)/i$.
When reduced to equal helicities $(\lambda_1,\lambda_2)=(\lambda_3,\lambda_4)$, \eq{hankelcondition} implies that $\M^\dagger\M$ in  non-forward scattering must be smaller than in forward one (generalising \cite{Bellazzini:2021oaj,Guerrieri:2022sod} to all helicities). When limited to the forward limit, instead, 
it implies that scattering of inelastic helicity must be suppressed w.r.t. the elastic one.

Since much of our understanding of dispersion relations relies on elastic scattering, \eq{hankelcondition} provides an intuitive way of readily extending previous results to inelastic scattering.}
    \\

\begin{figure}[htb]
   \centering
  \includegraphics[width=0.4 \textwidth]{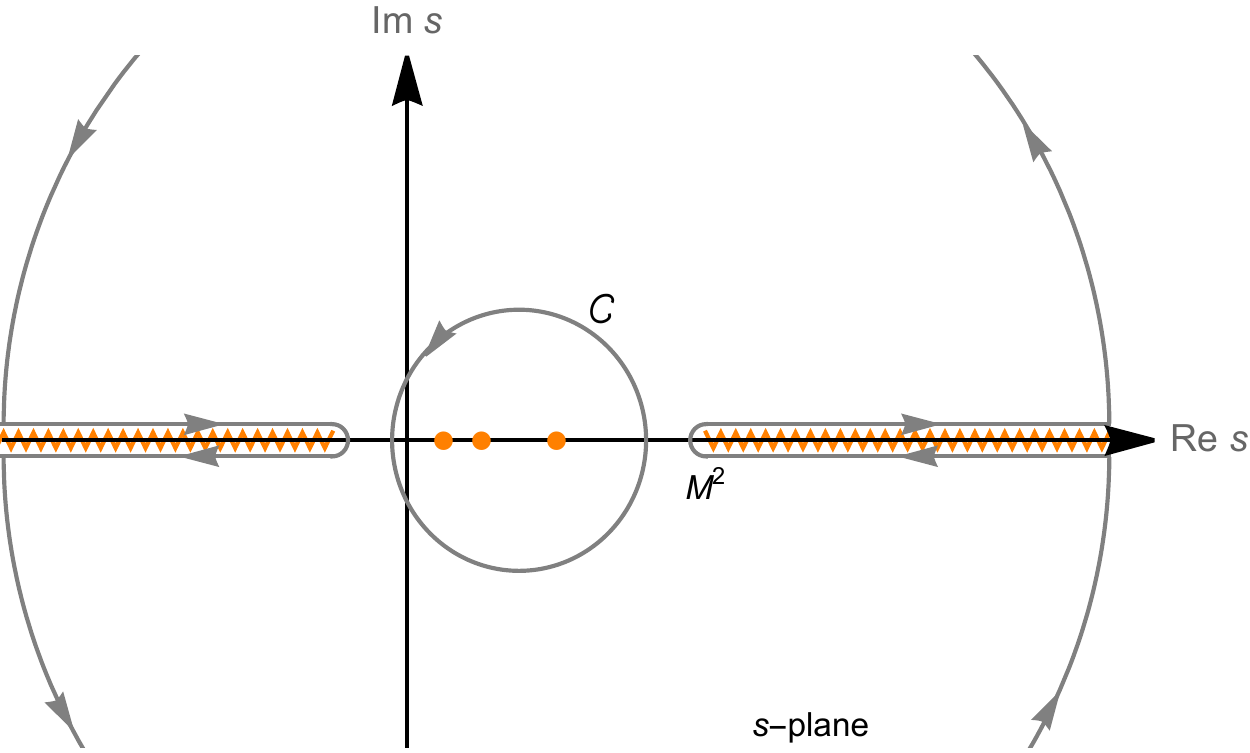}  
\caption{  \label{fig:contourArc} \footnotesize
Analytic structure of elastic-helicity amplitudes:  IR poles and subtraction points are schematically represented by orange dots, and  are well within the contour $\mathcal{C}$.   UV branch cuts are explicitly displayed, whereas IR ones are omitted. }
\end{figure}

\noindent
{\it ii)  Causality/Analyticity}\\
The center of mass scattering matrix elements 
\begin{equation}
\M_{\lambda_1 \lambda_2}^{\lambda_{3}\lambda_4}(s,t) \equiv   \langle 3^{\lambda_3} 4^{\lambda_4}| \M |1^{\lambda_1} 2^{\lambda_2}\rangle \quad \mbox{(c.o.m.)} 
\end{equation} 
are analytic functions in the complex  (Mandelstam) $s$ plane at fixed $t\leq 0$, except for branch cuts and poles located on the real axis
(in fact, our results rely only on  analyticity for large enough $|s|>R(t)$ with  $|R(t)|< M^2$ at small enough $|t|<M^2$, as proven in Ref.~\cite{Bros:1964iho}). 
These discontinuities are associated with physical thresholds (intermediate states exchanged in the $s$ or $u$ channels) as well as  kinematic singularities.
The latter,  classified long ago~\cite{Cohen-Tannoudji:1968lnm},  originate  either from the fact that helicity states are ill-defined when the momenta vanish, at $s=4m^2$ in the c.o.m frame,  or from angular-momentum selection rules.\footnote{\label{ftnt2}By rotational invariance, in  forward and backward scattering, the amplitude must behave like $\sqrt{-t/(s-4m^2)}^{|\lambda_{12}-\lambda_{34}|}$ and $\sqrt{-u/(s-4m^2)}^{|\lambda_{12}+\lambda_{34}|}$, respectively, where $\lambda_{ij}\equiv\lambda_i-\lambda_j$. For $m=0$ these are the only kinematic singularities  and are encoded in  little-group scaling factors. }  
{For instance}, the elastic-helicity scattering amplitude 
\begin{equation}
\M_{\lambda_1 \lambda_2}(s,t) \equiv \M_{\lambda_1 \lambda_2}^{\lambda_{1}\lambda_2}(s,t)\,,
\end{equation}
in the theory of a massive spin-2 particle,  exhibits only simple dynamical poles at $s=m^2$ and $3m^2-t$, and a kinematic higher order pole at $s=4m^2$
, see Fig.~\ref{fig:contourArc}.
  \\
  
\noindent
{\it iii)  Crossing symmetry} \\
This is a simple relation between amplitudes as functions of momenta, and holds in generic reference frames. In the forward or massless limit, it involves exchanging any two legs of an amplitude. At finite momentum exchange and mass, however, an additional boost must be performed to bring back the amplitude into the c.o.m frame. In a crossing transformation that takes one particle in the in/out state into an anti-particle of the out/in state, the resulting Wigner rotations generically mix all  helicities,
\begin{equation}\label{crossedM}
\! \M_{\lambda_1 \lambda_2}^{\lambda_3 \lambda_4}(s,t)  \! = \! \! \!\sum_{ \lambda_i^\prime =-S}^{S}X_{\lambda_1 \lambda_2 \lambda_3 \lambda_4}^{\lambda_1^\prime \lambda_2^\prime \lambda_3^\prime \lambda_4^\prime}(s,t) \M_{\lambda_1^\prime \bar{\lambda}_4^\prime}^{\lambda_3^\prime \bar{\lambda}_2^\prime}(u,t) \,,
\end{equation} 
with $S$  the spin of the particle, and $X$  the crossing matrix---writable in terms of a  string of Wigner-d matrices \cite{Cohen-Tannoudji:1968lnm,deRham:2017zjm,Hebbar:2020ukp,Davighi:2021osh}, where for later convenience we have defined $\bar\lambda_i\equiv -\lambda_i$.

{In this work, we exploit the fact that   for large center of mass energy  $m^2,-t\ll s$, the structure of $X$ greatly simplifies,
\begin{equation}
X_{\lambda_1 \lambda_2 \lambda_3 \lambda_4}^{\lambda_1^\prime \lambda_2^\prime \lambda_3^\prime \lambda_4^\prime}(s,t)\propto \prod_{i} \left(\frac{\sqrt{|t|}m}{s}\right)^{|\lambda_i-\lambda_i^\prime|} \,,
\end{equation}
providing a $\sqrt{|t|}m/s$ suppression  for any helicity change from the $\lambda_1 \bar{\lambda}_4 \to \lambda_1 \bar{\lambda}_2$ (high-energy) configuration.
So, for elastic helicity,  \eq{crossedM} becomes,
\begin{align}
\label{approximateX}
\M&_{\lambda_1 \lambda_2}(u,t) =   \M_{\lambda_{1} \bar{\lambda}_{2} }(s,t)  \\
\nonumber
 &+\left(c^{\lambda_1 \lambda_2}_{\lambda_1^\prime \lambda_2^\prime\lambda_3^\prime \lambda_4^\prime} \frac{\sqrt{|t|}m}{s}+O\left(\frac{tm^2}{su}\right)\right)  \M_{\lambda_1^\prime \lambda_2^\prime}^{\lambda_3^\prime \lambda_4^\prime}(s,t) \,,
 \end{align}
where in the first term of the second line we sum over the 8 inelastic amplitudes ${\lambda_1^\prime \lambda_2^\prime\lambda_3^\prime \lambda_4^\prime}$ with only one $\pm1$ helicity change w.r.t $\lambda_1 \bar{\lambda}_2\lambda_1 \bar{\lambda}_2$. Moreover,  $c$'s in \eqref{approximateX} are all known and bounded, $c\leq \sqrt{6}$ for spin-2 particles.
Notice that some of the inelastic amplitudes on the r.h.s. of \eq{approximateX}  are further suppressed by powers of $t$ due to angular momentum conservation close to the forward limit, see footnote~\ref{ftnt2}. 
\\
 
\noindent
{\it iv) Hermitian Analyticity}\\
Amplitudes in the upper and lower $s$-plane are related by complex conjugation, e.g. 
\begin{equation}
\label{HermitianAnalyticity}
\M_{\lambda_1 \lambda_2}^{\lambda_3 \lambda_4}(s+i\epsilon,t)=\M_{\lambda_3 \lambda_4}^{\lambda_1 \lambda_2\, *}(s-i\epsilon,t)
\end{equation}
for $s$ and $t$ real.   \\

\noindent
{\it v) Polynomial Boundedness}\\
The amplitude at fixed $t\leq 0$ is polynomially bounded in $s$; in particular the Froissart-Martin bound \cite{Froissart:1961ux,Martin:1962rt} for a gapped theory implies that,
\begin{equation}
\label{eq:froissart}
\lim_{s\to\infty} \M_{\lambda_1 \lambda_2}(s,t)/s^2 \to 0\,.
\end{equation}   
A similar bound has been extended recently to massless gravity in various dimensions \cite{Haring:2022cyf}. \\

\noindent
{\it vi) EFT separation of scales}\\ 
We assume we are dealing with a relativistic EFT where $m\ll M$, so that one can systematically calculate amplitudes in the low energy window,
\begin{equation}
m\ll E\ll M\,,
\end{equation}
 to any desired accuracy, provided one works at sufficiently high loop order and includes operators of sufficiently large dimension.
 In the context of massive gravity,  at sufficiently small energy, the EFT  is well described by a Lagrangian comprised of the Einstein-Hilbert term and the dRGT potential. 

Furthermore, because the theory is weakly coupled all the way to the cutoff (see footnote \ref{ftnt1}), we  assume that it is possible to neglect  the effects of  IR loops; these can  systematically be taken into account; see Refs.~\cite{Bellazzini:2020cot,Arkani-Hamed:2020blm,Bellazzini:2021oaj,Riembau:2022yse,Chala:2021wpj,Li:2022aby}.
\\

Now  the goal is to show for what values of the ratio $m/M$  the above assumptions are  compatible with each other, in the context of dRGT.   
Because of the simple analytic structure and the simple behaviour under crossing discussed in {\it ii)} and {\it iii)} respectively, we focus on elastic-helicity amplitudes.
We introduce the integral,
 \begin{equation}
 \label{def:IRarcs}
\A_{\lambda_1 \lambda_2}(t) \!=  \frac{1}{2} \!\oint_{\mathcal{C}} \! \frac{ds}{2\pi i}\frac{\M_{\lambda_1 \lambda_2}(s,t)+\M_{\lambda_1 \bar{\lambda}_2}(s,t)}{(s-2m^2+t/2)^3} \,,
\end{equation}
along a contour $\mathcal{C}$ in $s\in\mathbb{C}$ running around the origin at $4m^2\ll|s|\ll M^2$, so that it avoids the amplitude poles while remaining within the region of validity of the EFT, as shown in Fig.~\ref{fig:contourArc}.  
Then, $\A_{\lambda_1 \lambda_2}$ can  be calculated explicitly in terms of the free parameters of the EFT: $c_3$ and $d_5$ in the case of dRGT.~\footnote{$\A_{\lambda_1 \lambda_2}$ is independent of the subtraction points since, in  dRGT, the leading amplitudes give $(\M_{\lambda_1 \lambda_2}(s,t)+\M_{\lambda_1 \bar{\lambda}_2}(s,t))\sim s^2$ at fixed $t$, as opposed to $s^3$, meaning that $\A_{\lambda_1 \lambda_2}$ can be computed as the  $s\to \infty$ residue of the EFT amplitudes themselves. }

   Because of analyticity \emph{ii)}, $\mathcal{C}$  can be deformed  to run along the branch cuts and a big circle at infinity, which vanishes due to   \eq{eq:froissart} in {\it v)}.  
  Hermitian analyticity  {\it iv)} puts   $\A_{\lambda_1\lambda_2}$ in the form of a dispersive integral of $(\M-\M^\dagger)/i $, and by crossing symmetry~{\it iii)} it can be rewritten as a single  integral  over the physical values of the Mandelstam variable $s$. 
 
   The EFT scale separation {\it vi)}, allows us to work~at $m^2\ll |t| \ll M^2$, so that  crossing symmetry within the integral in $|s|\geq M^2$ takes the simple approximate form \eqref{approximateX}. 
         Using unitarity  {\it i)} we rewrite $(\M-\M^\dagger)/i $ as in \eq{optical} to obtain  the following UV representation for $\A_{\lambda_1\lambda_2}$,
{\begin{align}
      \label{eq:disp1}
&\A_{\lambda_1 \lambda_2}(t) =  \int_{M^2}^{\infty} \frac{ds}{2\pi}\frac{1}{(s-2m^2+t/2)^3}  \\
\nonumber
&\! \times \! \left[
\langle 3^{\lambda_1} 4^{\lambda_2} | \M^\dagger\M|1^{\lambda_1}2^{\lambda_2}\rangle +    \langle 3^{\lambda_1}  4^{\bar{\lambda}_2} | \M^\dagger \M  |1^{\lambda_1} 2^{\bar{\lambda}_2} \rangle \right] \\
&\qquad\qquad\qquad\qquad\qquad\qquad\qquad+ \frac{\sqrt{|t|}m}{M^2} \Delta_{\lambda_1 \lambda_2}\,.%\left(\A(0)\right)
\nonumber
\end{align}
where $\Delta_{\lambda_1 \lambda_2}$ captures  departures from elastic crossing in \eq{approximateX}, for $s\geq M^2$, and is bounded by a known linear function of other $\A_{\lambda_i \lambda_j}(0)$, see appendix \ref{section:bounderror}.

 The positivity bounds follow directly from the UV representation of $\A_{\lambda_1\lambda_2}$ in \eq{eq:disp1}.
In the forward limit  $t\to 0$,  $|3^{\lambda} 4^{\lambda^\prime}\rangle\to |1^{\lambda} 2^{\lambda^\prime}\rangle$ 
so that $\A_{\lambda_1 \lambda_2}(t\to 0)$  is a sum of squares, implying,
\begin{equation}
\A_{\lambda_1 \lambda_2}(0) \geq 0 \,,
\end{equation}
 with the equal sign obtained only in the free theory.  
For $t\neq 0$ instead, we use the fact that the matrix elements of $\M^\dagger \M$ are smaller than those at $t=0$, see \eq{hankelcondition}, 
and  obtain,
 \begin{equation}
\label{BoundFiniteT}
\frac{ \left|\A_{\lambda_1 \lambda_2}(t) \right|}{\A_{\lambda_1 \lambda_2}(0)}  \leq 1+O\left(\frac{\sqrt{|t|}m}{M^2}\right)\,. 
 \end{equation}
The  term $O(\sqrt{|t|}m/M^2)$ stems from $\Delta_{\lambda_1 \lambda_2}$ in \eq{eq:disp1} and, as discussed in appendix~\ref{section:bounderror}, is  bounded 
by the sum of the 8 known IR terms,  $(\!{\sqrt{|t|}m}/{M^2})\bar c^{\lambda_1 \lambda_2}_{\lambda_1^\prime \lambda_2^\prime\lambda_3^\prime \lambda_4^\prime} \!({\A_{\lambda^\prime_1 \lambda^\prime_2}\!(0)\!+\!\A_{\lambda^\prime_3 \lambda^\prime_4}\!(0)}\!)/2{\A_{\lambda_1 \lambda_2}\!(0)}$, summed as described below \eq{approximateX}, and where we defined $\bar c^{\lambda_1 \lambda_2}_{\lambda_1^\prime \lambda_2^\prime\lambda_3^\prime \lambda_4^\prime} \equiv |c^{\lambda_1\lambda_2}_{\lambda_1^\prime \lambda_2^\prime \lambda_3^\prime \lambda_4^\prime}+c^{\lambda_1\bar{\lambda}_2}_{\lambda_1^\prime \lambda_2^\prime \lambda_3^\prime \lambda_4^\prime}|/2$.  In this way, \eq{BoundFiniteT} can
be used  to formulate positivity bounds with complete control of terms  of order~$\sqrt{|t|}m/M^2$.

We remark that 
in the general case of scattering identical {massless} particles of arbitrary spin, we can write the exact inequality,
\begin{equation}
\label{BoundFiniteTMassless}
\frac{\left|\A_{\lambda_1 \lambda_2}(t) \right| }{\A_{\lambda_1 \lambda_2}(0)} \leq \left(1+t/2M^2\right)^{-3}\qquad (m=0)\,,
 \end{equation} 
 similarly to the massless scalar case of Ref.~\cite{Bellazzini:2021oaj}.

 \vspace{3mm}
 
The problem of finding all positivity constraints for massive spin-2 particles is quite complex, since crossing symmetry mixes hundreds of different amplitudes with each other, producing a nested network of positivity relations. These can in principle be solved with the methods of e.g. \cite{Caron-Huot:2020cmc,Caron-Huot:2021rmr,Tolley:2020gtv,Chiang:2021ziz,Bellazzini:2020cot,Bellazzini:2021oaj,Arkani-Hamed:2020blm},  but the advantage of working at leading order in $\sqrt{|t|}m/M^2$ is captured by the simplicity of \eq{BoundFiniteT}, which singles out   6 independent inequalities for the elastic helicities {$1^{\lambda_1}2^{\lambda_2}=1^02^0,1^02^+,1^+2^+,1^{=\!\!\!|\,\,}2^{=\!\!\!|\,\,},1^{=\!\!\!|\,\,}2^{0},1^{=\!\!\!|\,\,}2^{+}$}  (where we denote  by 0, +, and $=\!\!\!\!\!\!|\;$, the longitudinal, transverse and transverse-transverse helicities, with other elastic configurations  related to these ones by accidental parity, time-reversal and crossing in dRGT).
 The inequalities in \eq{BoundFiniteT}, via the IR representation \eq{def:IRarcs}, will be
 sufficient to constrain the parameter space of dRGT in the next section.  

\section{Positivity in dRGT}
\label{sec:positivitydRGT}

Scattering amplitudes in dRGT massive gravity are suppressed by  $m^2$ in the forward limit and, for some helicities,   grow rapidly at large~$|t|$. For $|t|\gg m^2$ this behaviour is incompatible with~\eq{BoundFiniteT}, for $|t|/M^2$ small enough. 

Of the six \ elastic-helicity configurations at our disposal,
the strongest bounds will come from~$\lambda_1\lambda_2=00,0+,++$, which give
 (from the dRGT action reported in Appendix~\ref{dRGTEFT}),
\begin{align}
\label{A00HighT}
\A_{00} \xrightarrow[m^2 \ll |t|]{} & \frac{t}{6\Lambda_3^6}\left(1 - 4 c_3 + 36 c_3^2 + 64 d_5\right) \\ 
\label{A0+HighT}
\A_{0+} \xrightarrow[m^2 \ll |t|]{} & \frac{t}{96\Lambda_3^6}\left(1 + 24 c_3 + 144 c_3^2 + 384 d_5\right)\\
\label{A++HighT}
\A_{+ +} \xrightarrow[m^2 \ll |t|]{} &  \frac{9t}{64\Lambda_3^6}\left(1-4c_3\right)^2\,,
\end{align}
while amplitudes involving the transverse polarisations do not grow with $|t|$.
This has to be contrasted with the values in the forward limit,
\begin{align}
\A_{00} \xrightarrow[t=0]{} & \frac{2m^2}{9\Lambda_3^6}\left(7 - 6 c_3 - 18 c_3^2 + 48 d_5\right) \\ 
\A_{0+} \xrightarrow[t=0]{} & \frac{m^2}{48\Lambda_3^6}\left(91 - 312 c_3 + 432 c_3^2 + 384 d_5\right)\\
\A_{+ +} \xrightarrow[t=0]{} &  \frac{m^2}{8\Lambda_3^6}\left(7 - 24 c_3^2 + 48 d_5\right) \,. \label{at0++}
\end{align}

Now, an EFT  with a large  range of validity can, by definition, be used at energies much larger than the particle mass, $m^2\ll | t|\ll M^2$. In this limit, the bounds from applying \eq{BoundFiniteT}  to (\ref{A00HighT}-\ref{at0++}), would converge to three lines in the $(c_3,d_5)$ plane, corresponding to the vanishing 
 of \eqref{A00HighT}, \eqref{A0+HighT} and \eqref{A++HighT}. These three lines have no common intersection, as illustrated in the left panel of Fig.~\ref{fig:exclusion1}.
 
\begin{figure*}[t]\centering
\includegraphics[height=6.9cm]{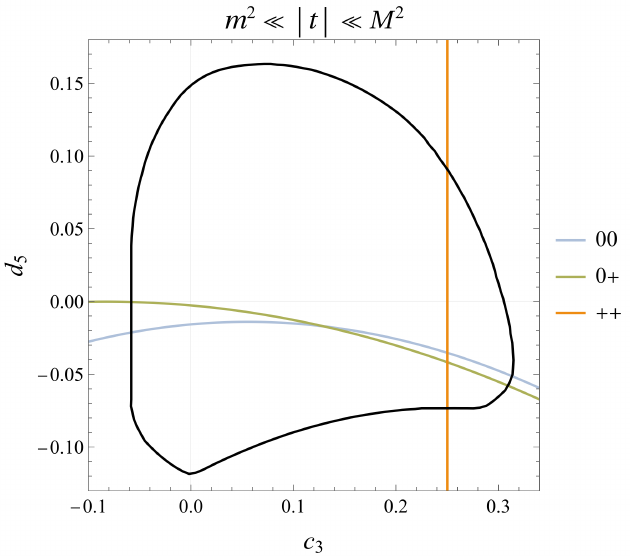}\hspace{0.5cm}
  \includegraphics[height=6.9cm]{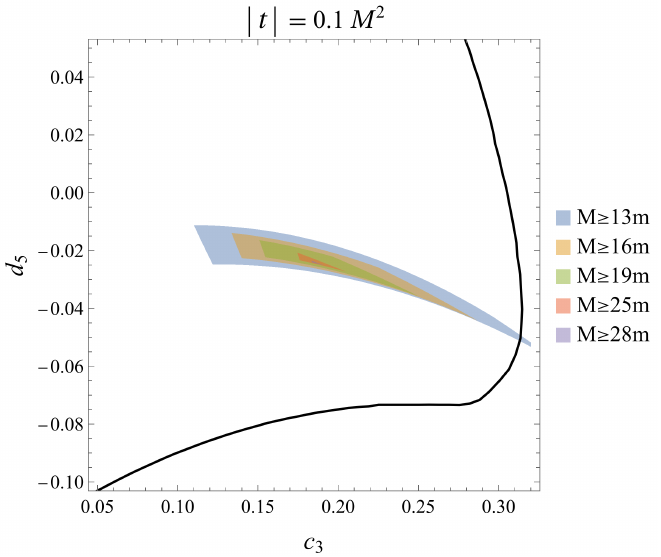}  
\caption{\footnotesize
 \it \label{fig:exclusion1} 
The  $(c_3,d_5)$ parameter space of dRGT massive gravity, and 
a comparison with the  forward-only positivity bounds from Ref.~\cite{Cheung:2016yqr} which carve the region  inside the closed black line.
LEFT:      In the $|t|\gg m^2$ limit, each elastic helicity reduces the parameter space to a line (corresponding to the vanishing of  \eqref{A00HighT} in blue, \eqref{A0+HighT} in green, and \eqref{A++HighT} in orange). In this limit, the lines do not intersect, and the theory is ruled out.
RIGHT: 
A close-up of the same figure, for finite values of  $|t|/m^2$ (we have used the exact amplitudes, rather than the ones expanded at large $|t|/m$ reported in the main text).
 Different shadings correspond to the allowed parameter space for different values of $M/m$, represented at fixed $|t|/M^2=0.1$. As the ratio between the cutoff and the mass increases, the parameter space shrinks and eventually disappears, hence providing \eq{mainBound}. 
}
\end{figure*}

 This implies that in dRGT massive gravity,  the cutoff of the theory cannot be arbitrarily large compared to the mass.
To quantify this, we run a bootstrap algorithm for the ratio $m^2/M^2$, assuming only the existence of a range $|t |\ll M^2$ for which dRGT is  a valid description of massive spin-2 scattering.  
For each value of $m^2/M^2$, we determine the set of points $(c_3,d_5)$ that 
are compatible with the finite-$t$ bound in \eq{BoundFiniteT};  if the set is not empty then we lower $m^2/M^2$ and repeat; if the set is empty, the value  is inconsistent with the assumptions {\it i)-vi)} and is discarded. 
The results of this algorithm are illustrated  in the right panel of  Fig.~\ref{fig:exclusion1}.

In this way, we find that the cutoff scales linearly with the mass, and is limited to being parametrically close to it,
\begin{equation}
\label{mainBound}
M  \leq  30\,\,  m  \times \left(\frac{0.1}{-t/M^2}\right)^{1/2} \,. 
\end{equation}
We have presented the bound in this way to highlight the fact that
it becomes stronger as the theory is evaluated at larger energies $|t|/M^2$, closer and closer to the cutoff, while 
 still being described by dRGT.
Since  $m^2/M^2$ is small, for $-t/M^2\leq 0.1$ it implies $m\sqrt{-t}/M^2\leq 0.01$  and we have checked that the error described below \eq{BoundFiniteT}  is indeed negligible.% and 

\section{Beyond dRGT: Higher Orders in the Energy expansion}
\label{secBDRGT}

In the previous section we have assumed that dRGT accurately describes massive spin-2 scattering  within the EFT.
In general, there might be terms of higher order in the energy expansion, beyond those of dRGT, that also contribute to the scattering amplitudes via terms with more powers of the energy. These enter $\A_{\lambda_1 \lambda_2}(t)$ as higher powers in~$t$. As long as these terms are suppressed by powers of $M$, and are controlled by  coefficients $\sim O(1)$ w.r.t dRGT, our arguments are modified only by higher powers of the small ratio~$|t|/M^2$. 
In this section, we relax this assumption, and study the possibility that above some intermediate scale $ E_*$, with $m \ll E_*<M$, dRGT transitions into a different theory, controlled by the most general higher-derivative EFT. Can such a theory exist?

If such a theory is dominated by large coefficients in just a few terms of higher-order  in the energy expansion,   we could apply the same arguments as in section~\ref{sec:positivitydRGT} to $\A_{\lambda_1 \lambda_2}(E_*^2 \ll |t|\ll M^2)/\A_{\lambda_1 \lambda_2}(0)$ (and to the generalisations of $\A_{\lambda_1 \lambda_2}(t)$ to more subtractions)  and exclude it, since a fixed order polynomial in $t/E_*^2$ would quickly exceed $1$ in~\eqref{BoundFiniteT}.

\paragraph{\bf Detuned potential.}
An example of such a theory is provided by detuning the graviton potential from its dRGT values, i.e. of all the dRGT relations in Appendix \ref{dRGTEFT} we only keep  the Fierz-Pauli mass tuning $b_1=-b_2$. Then the 2-to-2 amplitudes depend on 4 parameters, $(c_1,c_2,d_1,d_3)$, which we can constrain as follows.

The amplitude for $\lambda_1 \lambda_2=00$ at 0-th order in the mass grows as $\propto (c_2+3c_1/2 -1/4)s^4t$. We can therefore define the analog of \eq{def:IRarcs} with 4 subtractions $\A_{\lambda_1 \lambda_2}^{(4)}$ (i.e. exponent 5 at denominator), that also satisfies \eq{BoundFiniteT}, and
 implies at $|t|\gg m^2$ that  $c_2=-3c_1/2 +1/4+O(m^2/M^2)$. 
 Moreover,  at $t=0$, $\A_{\lambda_1 \lambda_2}\geq \A_{\lambda_1 \lambda_2}^{(4)} (M^2-2m^2)^2$, see Ref.~\cite{Bellazzini:2020cot}, implying $d_3=-d_1/2+3/32+O(m^2/M^2)$ (after imposing the $c_2$ tuning). These are the 
 dRGT relations spanned by $(c_1,d_1)$ and deliver the $\Lambda_3$-theory, reproducing Ref.~\cite{deRham:2018qqo}
and further showing that detuned massive gravity is also incompatible with positivity bounds as soon as $m^2$ and $M^2$ are parametrically separated.\footnote{\label{footdetuned} Further, using $\A_{\lambda_1 \lambda_2}$ with $\lambda_1 \lambda_2= 00$, $0+$, $++$ reduces this theory parameter space to three lines, $128d_1 - 43 + 36 c_1 - 108 c_1^2=0$, $16d_1 - 6 c_1 - 9 c_1^2=0$ and $c_1=1$, up to $O(m^2/M^2)$. Interestingly, including finite $m^2/M^2$ effects, we find that  in detuned massive gravity the bound on $M/m$ is exactly as in dRGT: $M \lesssim 30 m \sqrt{0.1 M^2/|t|}$, with error controlled by the chosen ratio $|t|/M^2$.}

\paragraph{\bf General higher derivative terms.}
Rather than a normal EFT with a finite number of leading higher-energy terms, we can even address the case of a theory with a tower of  infinitely many higher derivative terms with large coefficients, arranged such that their contributions to $\A_{\lambda_1\lambda_2}$ resum to a small function of $t/M$.
To answer this question we provide an alternative derivation of the  bound that led to the non-intersecting lines in the left panel of Fig.~\ref{fig:exclusion1} (for which we used $|t|\gg m^2$ in section \ref{sec:positivitydRGT}). In this derivation we will not assume that  $\A_{\lambda_1 \lambda_2}$ is at most linear in~$t$, as in dRGT, but allow for arbitrary powers of $t$ with arbitrary coefficients. On the other hand, in this derivation, we will work at zeroth order in $m$, keeping $\Lambda_3$ fixed (this is known as the \emph{decoupling limit}, in which the transverse polarizations decouple).

At this order, besides the simplification of crossing symmetry discussed in {\it iii)}, the EFT amplitudes also simplify 
 because the theory effectively reduces to that  of a massless shift-symmetric scalar,  a  photon, and  a graviton.   At high energy  we are thus able to write all-orders Ans\"atze, the relevant ones being,
 \begin{eqnarray}
\langle 3^0 4^0|\M|1^0 2^0\rangle  &= &  H(s,t) \label{h00},\\ 
\langle 3^+ 4^-|\M|1^+ 2^-\rangle&= & \langle 32\rangle^2   [14]^2 G_{+-}(s,t),\\
\langle 3^0 4^+|\M|1^0 2^+\rangle &= & \langle 4 1\rangle^2 [1  2]^2  G_{0+}(s,t) \label{h001},
\end{eqnarray}
where we have factored out little group scalings, and $H$ and $G_{\lambda_1 \lambda_2}$ are functions that contain only dynamical singularities.  Moreover, within the decoupling limit and within the EFT range of validity, they are also analytic functions,  since none of the 3-pt functions between one neutral Goldstone boson and the gauge boson give rise to on-shell poles.   
Crossing symmetry implies that $G_{0+}(s,t)=G_{0+}(u,t)$, $ G_{+-}(s,t)=G_{+-}(t,s)$, while $H$ is fully $s-t-u$ crossing symmetric. 
Therefore, their most general tree-level low-energy expressions are,
\begin{align}
\label{Hlow}
\!& H(s,t) \!=  \!  h_0(s^2+t^2+u^2)/2 + h_{1}stu+\ldots   \\
\label{Gpmlow}
\!&G_{+-}(s,t) \!=  \! f_0 + f_1 (s+t)+ f_2 (s^2+t^2)+ \ldots \\
\label{G0plow}
\!&G_{0+}(s,t) \!=   \! g_0 + g_1 t + g_2 (s^2+u^2)+g_{2}^\prime su + \ldots    
\end{align}

Refs.~\cite{Bellazzini:2020cot,Caron-Huot:2020cmc,Tolley:2020gtv,Caron-Huot:2021rmr} have derived
 bounds for all the coefficients in the most general EFT for scalars. These can be readily applied to the  ratios $h_i/h_0$, constraining them from above and below in appropriate units of $M$, independently of the value of all the other coefficients.
Similarly, Ref.~\cite{Henriksson:2021ymi,Henriksson:2022oeu,Haring:2022sdp} derived two-sided bounds for spin-1 particles, which can be read in terms of  $f_i/f_0$. 

In Appendix~\ref{section:moments} we perform a similar analysis, for amplitudes involving both spin-1 and spin-0 particles.
We exploit the fact that, again because of crossing symmetry, the  form factors  in Eqs.~(\ref{h00}--\ref{h001}) also control other amplitudes, namely, $\langle 3^- 4^+|\M|1^0 2^0\rangle   =   \langle 4 1\rangle^2 [1  3]^2  G_{0+}(t,s)$ and $\langle 3^+ 4^+|\M|1^+ 2^+\rangle  = \langle 3 4\rangle^2  [12]^2 G_{+-}(u,t)$. This allows us to study inelastic channels to  find lower and upper bounds on the $g_i$'s. 

A simple---albeit non-optimal---subset of these bounds reads, 
 \begin{align}
 \label{2sidedH}
-8 h_0& \leq   h_{1}M^2 \leq   \frac{3}{2}h_0, &  \\
 \label{2sidedF}
 - f_0& \leq   f_{1}M^2   \leq   f_0 ,& \\
   \label{2sidedG}
   -\frac{5}{2}g_0 &\leq  g_{1}M^2  \leq \frac{1}{3}\left(10 g_0 + 4 h_0 +7 f_0\right) \,, & 
\end{align}
and holds regardless of higher derivative terms, which are also similarly bounded. 

In a theory that reduces  to dRGT at low energies, and departs from it only by higher derivative terms, the most relevant terms $h_{0,1}$, $f_{0,1}$ and $g_{0,1}$ must match with dRGT, i.e.
\begin{equation}
\!\!\! -h_{1} t =\mbox{Eq.}~\eqref{A00HighT},\,\, \,    g_{1} t=\mbox{Eq.}~\eqref{A0+HighT} ,\,\,\,    \frac{3}{2}f_{1} t=\mbox{Eq.}~\eqref{A++HighT}.\nonumber
\end{equation}
The coefficients $h_0$, $g_0$, and $f_0$ are mass-suppressed and thus vanish at the order $O(m^0)$ that we assume in this section.
Therefore, combining these explicit expressions with the bounds in Eqs.~(\ref{2sidedH}-\ref{2sidedG})  leads to exactly the same situation as in the left panel of Fig.~\ref{fig:exclusion1}, but this time, independently of all higher derivative terms. This holds also for detuned massive gravity with higher energy terms, here with different expressions for $h_1,g_1,f_1$ in terms of $(c_1,d_1)$.

\paragraph{\bf Non-gravitational spin-2 theories.}

It is interesting to extend  our bounds to  non-gravitational  massive spin-2 theories, i.e. theories without  diffeomorphism-invariance, that would describe the dynamics of e.g. glueballs in gauge  theories.
One practical way to define these theories is by starting with a gravitational theory such as massive gravity with a generic potential (but keeping the Fierz-Pauli mass tuning) and then add diffeomorphism hard-breaking terms. The lowest-dimensional hard-breaking interaction for a parity even massive spin-2 particle corresponds to a single on-shell 3-point function with 2 momentum insertions, schematically $\partial^2 h^3$, different from the one in the Einstein-Hilbert term.  This new vertex can be chosen to be~\cite{Bonifacio:2018vzv, Hinterbichler:2013eza,Hinterbichler:2017qyt},
\begin{equation}
\delta\mathcal{L}=\zeta \frac{\mpl^2}{2}\epsilon^{\mu\nu\rho\sigma}\epsilon^{\alpha\beta\gamma\delta}h_{\sigma\delta} \partial_{\rho}h_{\mu\alpha}\partial_\gamma h_{\nu\beta}
\end{equation}
(parity odd spin-2 particles admit another trilinear with an $\epsilon$ tensor \cite{Bonifacio:2018vzv}).  Restricting to parity-even for simplicity,  the new trilinear  dramatically changes the $m\ll E \ll M$ behaviour of scattering amplitudes\footnote{An interesting limit is obtained by decoupling the Einstein-Hilbert term, $\mpl\to \infty$, while keeping the other interactions $\zeta/\mpl$, $c_i/\mpl$ and $d_i/\mpl^2$ finite, and where the strong-coupling scale of longitudinal modes is raised to $\Lambda_3$ by performing the aforementioned limit of the dRGT-tunings. In this corner of parameter space the theory becomes invariant under linear diffeomorphisms \cite{Hinterbichler:2013eza}, and its (in)consistency has been already studied in Ref.~\cite{Bonifacio:2016wcb}. }. In particular for the transverse-transverse polarizations (denoted~$\,\,|\!\!\!\!\!=$ and $=$) the leading energy behavior $m\ll E \ll M$ reads e.g. 
\begin{align}
\!\! \M_{=\!\!\!|\,\,=\!\!\!|\,\,}=-\frac{8}{3}\zeta^2 \frac{s^3}{\Lambda_3^6}  \,, \,\,\,\,
\M_{=\!\!\!|\,\,=}=\frac{8}{3}\zeta^2 \frac{(s+t)^3}{\Lambda_3^6}  %\Lambda_3^6=\mpl^2 m^4
\end{align}
 and likewise for the mixed transverse-longitudinal, e.g. $\M_{=\!\!\!|\,\, +}=-s^2(s+t)\zeta^2/\Lambda_3^6$,  or for only longitudinal modes, $\M_{0\, +}=\zeta s t^2(s+t)/6\Lambda_3^6m^2$, \ldots,  (after the scalar-sector tuning as required by positivity $c_2=-3c_1/2+1/4+o(m^2/M^2)$ and $d_3=3/32 - d_1/2 +o(m^2/M^2)$).

%These results are of course expected  since approximate diffeomorphisms is  the way to enforce a better $E\gg m$ behavior. 

Scattering with such a fast leading-energy scaling is in fact inconsistent with \eqref{BoundFiniteT} (and as well as with the results of Ref.~\cite{Caron-Huot:2022ugt}) unless $\zeta$ is strongly suppressed or the cutoff is parametrically close to the mass.  Indeed, for $|t|\gg m^2$ we have, 
\begin{align}
\A_{=\!\!\!| \,\, =\!\!\!|} =4t\frac{\zeta^2}{\Lambda_3^6}\,, \quad \A_{=\!\!\!|\,\,-}=\frac{ t}{2}\frac{\zeta^2}{\Lambda_3^6}\,,\quad \A_{0=\!\!\!|\,\,}= \frac{2t}{3}\frac{\zeta(1-4\zeta)}{\Lambda_3^6}\nn
\end{align}
 whereas the values in the forward limits are mass-suppressed: after enforcing the scalar-sector tuning,  $\A_{=\!\!\!| \,\, =\!\!\!|}(0) =(1-8c_1\zeta+32\zeta^2) m^2/\Lambda_3^6$, $\A_{=\!\!\!|\,\,-}=(4-4\zeta-6\zeta^2+c_1(-3+6\zeta)m^2/\Lambda_3^6$ and $\A_{0=\!\!\!|\,\,}=(1-4\zeta)(4-3c_1+8\zeta)2m^2/3\Lambda_3^6$. Analogously for the scalar-vector, where $\A_{0+}(|t|\gg m^2)=t^2\zeta/6\Lambda_3^6 m^2$ whereas the forward arc  $\A_{0+}(0)$ is mass suppressed, thus demanding again $|\zeta|\ll 1$.

However, as soon as $|\zeta|\ll 1$, e.g. by setting $\zeta=\bar{\zeta}m^2/M^2$ for a finite $\bar{\zeta}$,  this new coupling essentially drops out from the constraints \eqref{BoundFiniteT} for  the longitudinal-only polarizations $\lambda_1 \lambda_2=00, ++$. It survives  in the $0+$ but only proportionally to a $t/M^2$ factor which is of the same order of other higher derivative corrections, hence irrelevant. Indeed, the bounds \eqref{2sidedH},  \eqref{2sidedF}, and \eqref{2sidedG} still apply on the $h_1$, $f_1$, and $g_1$, regardless of the higher orders in $t$ that affect instead $(h,\,f,\,g)_{i\geq 2}$. All in all, for such a small $\zeta$, we simply recover the same gravitational constraints for the longitudinal modes that we have already studied. See them explicitly in footnote~\ref{footdetuned}. \\

%\vspace{2mm}
In conclusion,  theories of massive spin-2 particles cannot have a parametric separation of scales $M/m$, independently of how they are modified at high energy. 
 
\section{Conclusions and Outlook}
\label{sec:conclusions}

The EFT of massless gravitons is a priori consistent  from the smallest energy scale in the universe $H_0\sim 10^{-42}$ GeV, to the largest one $\mpl\sim 10^{18}$ GeV, i.e. over about  60 orders of magnitude. 
The results presented in this paper show that consistency of the EFT of a massive graviton  is instead confined into a narrow energy window, spanning  from the graviton mass by at most one order of magnitude.  This constitutes an improvement of 15 orders of magnitude w.r.t.  previous bounds.

We devised new and simple positivity bounds based on an approximate crossing symmetry that is valid in weakly coupled EFTs with a hierarchy between mass $m$ and cutoff $M$.
The simple relations we obtain can be employed within dispersion relations (based on unitarity and causality) to study complex problems, such as massive higher-spin scattering. They lead to \eq{BoundFiniteT}, which bounds the energy growth of elastic-helicity  amplitudes to lie within a certain envelop. 
With this, we found that massive gravity can not  sustain a parametrically large mass hierarchy,   see \eq{mainBound} and Fig.~\ref{fig:exclusion1}, as it would fail our positivity bounds.
This conclusion is robust w.r.t. the inclusion of arbitrary  number of higher derivative terms, as well as  higher order corrections to our version of simplified crossing symmetry. A similar conclusion holds for non-gravitational massive spin-2 particles. 
  
{While our results exclude massive gravity with just the graviton and nothing else in the spectrum below $O(10)m$, they do not}  exclude theories with no parametrically large separation of scales, such as KK gravitons that arise from the compactification  of extra dimensions,   or theories that do not fulfil our assumptions {\it i)-vi)}. 
Moreover, the quantitative bound in \eq{mainBound} becomes inaccurate if one pushes it to the regime  $m^2\sim t \sim M^2$, where it seems to become stronger.
  In the context of gravity, however, more stringent bounds  would be  incompatible with the inherently flat-space formulation of the dispersive approach,
  as curvature corrections can no longer be neglected for~$M\sim m \sim H_0$.  The extension of positivity bounds to theories in non-flat backgrounds is very interesting \cite{Baumann:2015nta,Grall:2021xxm},  albeit rather subtle~\cite{Creminelli:2022onn}. 

 It would be interesting  to apply  the techniques developed in this paper to theories with massive higher spin $J\geq 3$ states, along the lines of  \cite{Bellazzini:2019bzh}.  It was shown there that, contrary to massive gravity~\cite{Cheung:2016yqr}, forward-only positivity bounds were already sufficient to rule out theories dominated by the most relevant interactions, but didn't exclude the possibility that an EFT with a parametric separation between $m$ and $M$ could exist, if dominated by  less relevant interactions. 
 The all-derivative order argument presented in section~\ref{secBDRGT} should be sufficient to close this door, and exclude any EFT for a single massive higher-spin particle.

Another interesting direction is to extend the analysis beyond our approximation  and derive  a version  of \eq{mainBound} that remains valid even for $m\sim M$~{\cite{inpreps2}}. 
While this is not motivated in the framework of massive gravity, as discussed above, it would teach us about the properties of spin-2 resonances, such as  glueballs in gauge theories and QCD.

\section*{Acknowledgments}

We are  grateful to Francesco Bertucci, Alfredo Glioti, Andrea Guerrieri, Aditya Hebbar, Brian McPeak, Javi Serra and Alessandro Vichi for useful discussions. 
We thank Cliff Cheung, Brian McPeak and Scott Melville for many constructive comments on the draft.  The work of G. I., S. R. and F. R. is supported by the Swiss National Science Foundation under grants no. 200021-205016 and PP00P2-206149.

\section*{Conventions}
We work in the $\eta_{\mu\nu}=\mathrm{diag}(-, +, +, +)$ metric signature. The Mandelstam invariants in the in-out convention are $-s=(p_1 + p_2)^2=(p_3 + p_4)^2$, $-t=(p_1 - p_3)^2=(p_2 - p_4)^2$, $-u=(p_1 - p_4)^2=(p_2 - p_3)^2$ with $s+t+u=4m^2$. The angle $|\cdot\rangle$ and square $|\cdot]$ brackets represent spinor-helicity variables. 
The scattering amplitudes $\M$ are defined by factoring out the momentum-conservation Dirac-delta function, i.e.  $S=\mathbb{I}+i T$ with  e.g. $\langle 3^{\lambda_3} 4^{\lambda_4}  |T |1^{\lambda_1} 2^{\lambda_{2}}\rangle=(2\pi)^4\delta^4(p_1+p_2-p_3-p_4) \langle 3^{\lambda_3} 4^{\lambda_4}  |\M |1^{\lambda_1} 2^{\lambda_{2}}\rangle$ for a 4-body interaction. The helicity indices inside in the partial waves states are treated as label of the states, so that $\langle \lambda_1 \lambda_2|$ is the ``bra'' vector dual to the ``ket'' $| \lambda_1 \lambda_2\rangle$, keeping i.e. the same ordering.

\appendix

\section{dRGT} 
\label{dRGTEFT}

For an on-shell formulation of massive gravity and dRGT see e.g.~\cite{Falkowski:2020mjq}.  
For the original off-shell formulation, consider the effective Lagrangian~\cite{deRham:2010ik,deRham:2010kj} 
\begin{equation}
S=  \int d^4x \sqrt{-g} \left[\frac{\mpl^2}{2} R - \frac{\mpl^2 m^2}{8}V(g,h)\right]
\end{equation}
where $V(g,h) =V_2 + V_3 + V_4$ is expressed in terms of an auxiliary metric $g_{\mu\nu} \equiv \eta_{\mu\nu}+h_{\mu\nu}$ as 
$V_2=  b_1 \langle h^2  \rangle + b_2 \langle h  \rangle^2$,  $V_3=c_1 \langle h^3  \rangle + c_2 \langle h^2 \rangle \langle h \rangle+ c_3 \langle h  \rangle^3 $, $V_4=d_1\langle  h^4 \rangle + d_2 \langle h^3 \rangle \langle h \rangle+ d_3 \langle h^2  \rangle^2 + d_4 \langle h^2  \rangle \langle h\rangle^2 +d_5 \langle h \rangle^4$,  with $\langle h\rangle\equiv h_{\mu\nu}g^{\mu\nu}$, $\langle h^2\rangle\equiv g^{\mu\nu}h_{\nu\rho}g^{\rho\sigma}h_{\sigma\mu}$, etc. The coefficients depend on $c_3$ and $d_5$, after imposing the ghost-free conditions $b_1 = 1=-b_2$, $c_1 = 2c_3 + 1/2$, $c_2 = -3c_3 -1/2$, $d_1= -6 d_5+3c_3/2 +5/16$, $d_2= 8 d_5 - 3 c_3/2 -1/4$, $d_3 = 3d_5- 3 c_3/4 - 1/16$, and $d_4 = -6 d_5+3 c_3/4$. These terms correspond to the leading effects in the energy expansion.

\section{Bounding the error}
\label{section:bounderror}
In this appendix we show how the error  in the last line of \eqref{eq:disp1} is bounded by known quantities. 
Using crossing, hermitian analyticity and renaming dummy indices,  it can be written as,
\begin{align}
\label{errorintegral}
\nonumber
\frac{\sqrt{|t|}m}{M^2} \Delta_{\lambda_1 \lambda_2}\equiv& \int_{M^2}^{\infty} \frac{ds/2\pi}{(s-2m^2+t/2)^3}\left[\delta X_{\lambda_1 \lambda_2 \lambda_1 \lambda_2}^{\lambda_1^\prime \bar{\lambda}_4^\prime \lambda_3^\prime \bar{\lambda}_2^\prime}\right.\nn \\\left.+\delta X_{\lambda_1 \bar{\lambda}_2 \lambda_1 \bar{\lambda}_2}^{\lambda_1^\prime \bar{\lambda}_4^\prime \lambda_3^\prime \bar{\lambda}_2^\prime}\right] 
&  \times \langle 3^{\lambda_3^\prime } 4^{\lambda_4^\prime} | \M^\dagger \M | 1^{\lambda_1^\prime} 2^{\lambda_2^\prime} \rangle 
\end{align} 
where, from \eq{approximateX}, $\delta X \equiv X(t)-X(0)$ inside \eqref{errorintegral} can be expanded at leading order in $m,-t\ll s$ as  $\delta X_{\lambda_1 \lambda_2 \lambda_1 \lambda_2}^{\lambda_1^\prime \bar{\lambda}_4^\prime \lambda_3^\prime \bar{\lambda}_2^\prime} = \sqrt{-t}\frac{m}{s} c_{\lambda_1^\prime \lambda_2^\prime \lambda_3^\prime \lambda_4^\prime}^{\lambda_1\lambda_2}+O\left({tm^2}/{M^4}\right)$.
Using \eq{hankelcondition} we can bound the discontinuities in the UV in terms of the elastic and forward ones,
\begin{align}
\nonumber
&2\left| \langle 3^{\lambda_3^\prime } 4^{\lambda_4^\prime} | \M^\dagger \M | 1^{\lambda_1^\prime} 2^{\lambda_2^\prime} \rangle  \right| \leq \langle 1^{\lambda_1^\prime } 2^{\lambda_2^\prime} | \M^\dagger \M | 1^{\lambda_1^\prime} 2^{\lambda_2^\prime} \rangle \\
\nonumber 
&+ \langle 1^{\lambda_1^\prime } 2^{\bar{\lambda}_2^\prime} | \M^\dagger \M | 1^{\lambda_1^\prime} 2^{\bar{\lambda}_2^\prime} \rangle + \langle 3^{\lambda_3^\prime } 4^{\lambda_4^\prime} | \M^\dagger \M | 3^{\lambda_3^\prime} 4^{\lambda_4^\prime} \rangle  \\
&+ \langle 3^{\lambda_3^\prime } 4^{\bar{\lambda}_4^\prime} | \M^\dagger \M | 3^{\lambda_3^\prime} 4^{\bar{\lambda}_4^\prime} \rangle  
\end{align}
that enter in  \eq{eq:disp1} at $t=0$. 
The  error \eqref{errorintegral} is thus bounded as 
\begin{align}
| \Delta_{\lambda_1 \lambda_2}| \leq    
\frac{(\A_{\lambda_1^\prime \lambda_2^\prime} (0)+ \A_{\lambda_3^\prime \lambda_4^\prime}(0)) |c^{\lambda_1\lambda_2}_{\lambda_1^\prime \lambda_2^\prime \lambda_3^\prime \lambda_4^\prime}+c^{\lambda_1\bar{\lambda}_2}_{\lambda_1^\prime \lambda_2^\prime \lambda_3^\prime \lambda_4^\prime}|}{4(1+\frac{t/2}{M^2-2m^2})^3}   \nn
\end{align}
to leading order in the expansion of $\delta X$, where we have also used that $s\geq M^2$ in \eqref{errorintegral}. 
Importantly, this is written in terms of the IR expression for $\A$ at $t=0$, which are calculable within the EFT, rather than in terms of unknown UV quantities.
 The expansion of $X$ is also known, and its coefficients are bounded, for instance $|c|\leq \sqrt{j(j+1)}$ with $j=2$  the spin of the scattered states. Moreover, only a finite number of them enter to any order in the expansion of $X$ -- see discussion below \eq{approximateX} -- and only up to 8 entries enter at leading order, as relevant for the~$c$'s.  Higher order contributions in the expansion of $\delta X$ are also bounded in this way.

\section{Two-sided bounds from Moments}
\label{section:moments}

In this section we prove the two-sided bounds \eqref{2sidedH}, \eqref{2sidedF} and \eqref{2sidedG}. 
Contrary to section~\ref{sec:positivitydRGT}, here we exploit the expansion of dispersion relations at $t\approx 0$, order by order in $t$. There are many implementations of this idea, that differ by how they extract information from the UV integrals: using positive geometry~\cite{Arkani-Hamed:2020blm}, semidefinite optimization~e.g.~\cite{Caron-Huot:2020cmc,Caron-Huot:2021rmr,Albert:2022oes,Fernandez:2022kzi},  or moment theory \cite{Bellazzini:2020cot,Chiang:2021ziz,Bellazzini:2021oaj}. Here we use the latter, which allows to easily derive analytic   bounds.

We define  $s$-channel dispersion relations for amplitudes stripped from their little group scalings [L.G.]
\begin{align}\label{arc_mom}
&_n\tilde{\mathcal{A}}_{\lambda_1\lambda_2}^{{\lambda_3\lambda_4}}(t)= \frac{1}{2i\pi} \!\oint_{\mathcal{C}} \! \frac{ds}{s^{n+3}}\frac{\mathcal{M}_{\lambda_1\lambda_2}^{\lambda_3\lambda_4}(s,t) }{\text{[L.G.]}} 
\,,
\end{align}
where [L.G.] is $\langle 32\rangle^2[14]^2/s^2$ for $(+-\to+-)$ and $  \langle 4 1\rangle^2 [1  2]^2/s^2$ for $(0+\to0+)$, \ldots,  while it's 1 for $(00\to00)$. 
 We first focus on elastic scattering $_n\tilde{\mathcal{A}}_{\lambda_1\lambda_2}^{{\lambda_1\lambda_2}}(t)\equiv \tilde{\mathcal{A}}_{\lambda_1\lambda_2}^n(t)$. 
The direct evaluation of \eq{arc_mom} provides an IR representation in terms of the Wilson coefficients defined in Eqs.~(\ref{Hlow}-\ref{G0plow}), e.g. 
\begin{equation}
\label{IRrepOfH}
 \tilde{\mathcal{A}}_{0 0}^n(t)= \left\{\begin{array}{lr}
 h_0-h_{1} t+ h_2  t^2+\ldots & n=0 \\
\frac{2}{3}h_2  t+\ldots & n=1 \\
\frac{1}{3} h_2 +\ldots  & n=2
\end{array}\right. 
\end{equation}
etc., while $ \tilde{\mathcal{A}}_{0 +}^0(t)=g_0+t g_1+\ldots $,  $ \tilde{\mathcal{A}}_{-+}^0(t)+ \tilde{\mathcal{A}}_{++}^0(t)= 2f_0+ f_1 t  +\ldots$,  and $\tilde{\mathcal{A}}_{-+}^0(t)- \tilde{\mathcal{A}}_{++}^0(t) =f_1 t+\ldots$, and so on.

The $ \tilde{\mathcal{A}}_{\lambda_1\lambda_2}^n$  admit also a UV representation, from  deforming the contour along the branch cuts.
Further expanding in partial waves, the $ \tilde{\mathcal{A}}_{00}^n$ and $ \tilde{\mathcal{A}}_{0+}^n$ take the form,
    \begin{align}\label{uvrepapp}
\!\!\sum_\ell 8(2\ell+1)  \int_{M^2}^\infty \!\!\!\! ds K^n_{\lambda_1\lambda_2}(s,t) P^{(0,2|\lambda_{12}|)}_{\ell-|\lambda_{12}|}(1+\frac{2t}{s}) \!\!
\end{align}
where  $P^{(a,b)}_\ell(x)$  are Jacobi polynomials and  the kernels $K^n_{\lambda_1\lambda_2}$  are given by,
\begin{align}
&K^n_{00}=\left(\frac{1}{s^{n+3}}+(-1)^n\frac{1}{(s+t)^{n+3}} \right)\mathrm{Im}\mathcal{M}_{00}^\ell\\
&K^n_{0+}=\left(\frac{1}{s^{n+1}}+(-1)^n\frac{1}{(s+t)^{n+1}}\right)\frac{\mathrm{Im}\mathcal{M}_{0+}^\ell }{s^2}\,. %\\
\end{align}
A similar but longer expression holds for  $\tilde{\mathcal{A}}_{-+}^0 \pm \tilde{\mathcal{A}}_{++}^0$. 

Expanding  in powers of $t$, defining $J^2 \equiv \ell(\ell+1)$ and using $P^{0,2\lambda}_{\ell-\lambda}(1+\epsilon)=1+\left[J^2-\lambda(1+\lambda))\right]\epsilon+O(\epsilon^2)$, we can write \eq{uvrepapp}  in terms of moments,
\begin{equation}\label{2Dmom}
\!\! \mu_{n, m}^{\lambda_i\lambda_j}=\sum_J{}^\prime \int_{M^2}^\infty \frac{ds}{s^{n+3}}J^{2m}\mathrm{Im} \M^{J^2}_{\lambda_i\lambda_j}(s)\geq 0\, 
\end{equation}
of the 2-dimensional positive measures $\mathrm{Im} \M^{J^2}_{\lambda_i\lambda_j}(s)$.\footnote{These are defined by the following partial wave expansion 
$\M_{\lambda_1 \lambda_2}^{\lambda_3\lambda_4}= 8\pi \sum_\ell (2\ell+1) d^{\ell}_{\lambda_{12} \lambda_{34}}(\theta) {\M^{\ell}}_{\lambda_1 \lambda_2}^{\lambda_3 \lambda_4}(s)$, with $ \M^{J^2}_{\lambda_i\lambda_j}(s) \equiv {\M^{\ell(J)}}_{\lambda_i \lambda_j}^{\lambda_i \lambda_j}(s)$ and $\sum_J{}^\prime=\sum_J 8\sqrt{1+4J^2}=\sum_\ell8 (2\ell+1)$ with $\ell \geq |\lambda_i -\lambda_j |$. We expand in partial waves for identical massive particles and take the massless limit afterwards -- this removes factors of $2$ from our expressions. }   
Matching powers of $t$ between the IR and UV representations Eqs.~(\ref{IRrepOfH}, \ref{uvrepapp}) we can write  Wilson coefficients in terms of  moments,
\begin{align}\label{Wilson_mom1}
&h_0=2\mu_{0,0}^{00}\, ,\quad &h_1=3\mu_{1,0}^{00}-2\mu_{1,1}^{00}\, ,\\ \label{Wilson_mom2}
&g_0=2\mu_{0,0}^{+0}\, ,\quad & g_1=-5\mu_{1,0}^{+0}+2\mu_{1,1}^{+0}\, ,\\ \label{Wilson_mom3}
&f_0=\mu_{0,0}^{-+}+\mu_{0,0}^{++}\, ,\quad & f_1=\mu_{1,0}^{-+}-\mu_{1,0}^{++}\, .
\end{align} 

Considering $\tilde{\mathcal{A}}^{n}_{\lambda_1\lambda_2}$ with $n>0$ reveals that Wilson coefficients admit more than one representation in terms of moments --- a consequence of crossing symmetry. For instance,  $h_2$ appears in $n=0$ as well as in $n=2$ of \eqref{IRrepOfH}, while $f_1$ appears in both $\tilde{\mathcal{A}}_{-+}^0\pm \tilde{\mathcal{A}}_{++}^0$. This leads to sum rules among moments, or  \emph{null constraints}~\cite{Caron-Huot:2020cmc,Tolley:2020gtv}. 
For our purpose it will be enough to use the simplest ones,  
\begin{align}\label{null_cons1}
8\mu_{2,1}^{00}&=\mu_{2,2}^{00}\, ,\\
7\mu_{1,0}^{-+}&=\mu_{1,1}^{-+}+\mu_{1,1}^{++}\, ,\nonumber
\end{align}
which connect moments in  $J^2$ to moments in  $1/s$.

All positivity relations satisfied by moments can be obtained by integrating positive polynomials in $J^2$ and $1/s$, see e.g.~\cite{Chiang:2021ziz,Bellazzini:2021oaj}.
From  positive monomials, it follows that all moments are positive, which directly leads to $g_0,h_0,f_0\geq 0$.
Instead, from the   polynomial $(1-M^2/s)$  (positive because the measure is supported for $s\geq M^2$), we find 
that moments in $1/s$ are monotonically decreasing,
\begin{equation}
\label{monotonicityMom}
\mu_{n,m}^{\lambda_i\lambda_j}\geq M^2\mu_{n+1,m}^{\lambda_i\lambda_j}\,. 
\end{equation}
Finally, from the positive quadratic polynomial $(a+b J^2/s)^2$, we find positive definiteness of the Hankel matrix,
\begin{align}
&\mathrm{det}\begin{pmatrix}
\mu_{0,0}^{\lambda_i\lambda_j} & \mu_{1,1}^{\lambda_i\lambda_j}\\
\mu_{1,1}^{\lambda_i\lambda_j}& \mu_{2,2}^{\lambda_i\lambda_j}
\end{pmatrix}>0\, .\label{Hankel_cond}
\end{align}
These positive relations, possibly supplemented by null constraints \eq{null_cons1}, lead to lower and upper bounds for all coefficients in units of the lowest ones.
Indeed, the conditions \eqref{monotonicityMom} and \eqref{Hankel_cond}  for $00$-scattering combined with \eqref{null_cons1} give $\mu_{1,1}^{00}M^2\leq 8\mu_{0,0}^{0,0}$, and therefore $M^2 h_1/h_0\leq 3M^2\mu_{1,0}^{00}/2\mu_{0,0}^{00}\leq 3/2$ and $ M^2 h_1/h_0 \geq -M^2\mu_{1,1}^{00}/\mu_{0,0}^{00}\geq -8$, proving \eq{2sidedH}.  Likewise, it is easy to prove
 \eqref{2sidedF} and the lower bound of \eqref{2sidedG}. Although not optimised, these relations are conservative.

For the upper bound of $g_1$ in \eqref{2sidedG} we must instead consider inelastic channels $\mathcal{M}_{00}^{-+}$ and $\mathcal{M}_{-+}^{00}$. As discussed above \eq{2sidedH},  these are controlled by the same function $G_{0+}$ in \eqref{G0plow}, and lead to another representation of $g_1$,
\begin{align}\nonumber
&g_1=\left.\frac{1}{2}\left(_1\tilde{\mathcal{A}}_{00}^{-+}+\,\,_1\tilde{\mathcal{A}}_{-+}^{00}\right)\right|_{t=0}
=-\frac{1}{4}\sum_J{}^\prime J^2\!\!\int_{M^2}^\infty \frac{ds}{s^4}\\    
\label{inel_g1} 
 &\Biggl[\frac{\sqrt{J^2-2}}{J}\left((\overline{\mathrm{Im}} \M_{00}^{-+})_{J^2}+(\overline{\mathrm{Im}} \M_{-+}^{00})_{J^2} \right) \\ 
\nonumber
&+\left( (\overline{\mathrm{Im}}\M_{0+}^{+0})_{J^2}+(\overline{\mathrm{Im}} \M_{+0}^{0+})_{J^2} \right) \Biggr]
\end{align}
where   
$(\overline{\mathrm{Im}} {\M}_{\lambda_1 \lambda_2}^{\lambda_3 \lambda_4})_{J^2} \equiv \langle \lambda_3 \lambda_4  | (\M^\dagger \M)_{\ell}| \lambda_1 \lambda_2 \rangle/2$.

Now, a bound on $g_1$ emerges  from inequalities between elastic and inelastic partial waves implied by unitarity.
{Positivity of the norm for the partial waves amplitudes $\M(|00\rangle +|-+\rangle)$ and $\M(|+0\rangle+|0+\rangle)$, analogous to \eq{hankelcondition}, 
with $(J^2-2)^{1/2}/J \leq 1$, puts \eq{inel_g1} in the form, }
\begin{align}
 g_1 \leq \sum_J{}^\prime  \frac{J^2}{4}\int_{M^2}^\infty \!\!\frac{ds}{s^4} \Biggl[ \mathrm{Im} \M_{00}^{J^2}+\mathrm{Im}\M_{-+}^{J^2} 
+ 2\mathrm{Im} \M_{+0}^{J^2} \Biggr]\,. \nonumber
\end{align}
Using \eqref{Wilson_mom2} and observing that $\sum_J{}^\prime$ in \eqref{inel_g1}  runs over a restricted set of $J$-values w.r.t.  \eqref{2Dmom}, gives,
\begin{equation}
\label{uppBg1}
6 \mu_{1,1}^{+0}\leq 20 \mu_{1,0}^{+0}+\mu_{1,1}^{00}+\mu_{1,1}^{-+}\, .
\end{equation}
This upper bound, together with   \eqref{Wilson_mom1}, \eqref{Wilson_mom2}, \eqref{Wilson_mom3},  the null constraints \eqref{null_cons1},  and the constraints  \eqref{monotonicityMom} and \eqref{Hankel_cond}, implies  \eqref{2sidedG}.  Repeating this analysis for higher moments one can bound higher order coefficients as well, e.g. $0\leq f_2 M^4\leq f_0$,  etc \ldots  %$0\leq (2g_{2}-g_2^\prime)M^4< g_0$,

In the decoupling limit discussed here, transverse modes are decoupled and, moreover, they have no impact on bounds.
Beyond this limit, at finite $m$, they can be included back in the analysis by extending the EFT analytic structure of the form factors to include their poles. 
Extra poles  are best addressed via the functional approach of \cite{Caron-Huot:2021rmr,Caron-Huot:2022ugt,Hong:2023zgm}, and produce relative corrections $O(m^2/M^2\log M^2/m^2)$, as estimated in the eikonal limit   \cite{Bellazzini:2022wzv} of the functionals.

\bibliography{bibs} 
\bibliographystyle{utphys}

\end{document}